# Chemically edge-carboxylated graphene enhances thermal conductivity of polyetherimide-graphene nanocomposites

*Fatema Tarannum[a], Rajmohan Muthaiah[a], Swapneel Danayat[a], Kayla Foley[b], Roshan S. Annam[a], Keisha B. Walters[b], Jivtesh Garg[a]\**

a School of Aerospace and Mechanical Engineering, University of Oklahoma, Norman, OK, USA 73019

b Department of Chemical Engineering, University of Arkansas, Fayetteville, AR, USA 72701

ABSTRACT: In this work, we demonstrate that edge-oxidation of graphene can enable larger enhancement in thermal conductivity ($k$) of graphene-nanoplatelet (GnP)/polyetherimide (PEI) composites relative to oxidation of the basal plane of graphene. Edge oxidation (EGO) offers the advantage of leaving the basal plane of graphene intact, preserving its high in-plane thermal conductivity ($k_{in}$ > 2000 Wm$^{-1}$K$^{-1}$), while, simultaneously, the oxygen groups introduced on graphene edge enhance interfacial thermal conductance through hydrogen bonding with oxygen groups of polyetherimide (PEI), enhancing overall polymer composite thermal conductivity. Edge oxidation is achieved in this work, by oxidizing graphene in presence of sodium chlorate and hydrogen peroxide, introducing carboxyl groups on the edge of graphene. Basal-plane oxidation of graphene (BGO), on the other hand, achieved through Hummers method, distorts $sp^2$ carbon-carbon network of graphene, dramatically lowering its intrinsic thermal conductivity, causing the BGO/PEI composite $k$ to be even lower than pristine GnP/PEI composite $k$ value. The resulting thermal conductivity of EGO/PEI composite is found to be enhanced by 18%, whereas that of BGO/PEI composite is diminished by 57%, with respect to pristine GnP/PEI composite for 10 weight% GnP content. 2-dimensional Raman mapping of graphene nanoplatelets is used to confirm and distinguish the location of oxygen functional groups on graphene. The superior effect of edge bonding presented in this work can lead to fundamentally novel pathways for achieving high thermal conductivity polymer composites.

## 1. INTRODUCTION

Thermal management has become a challenging issue in modern electronics due to continuous miniaturization of electronic components which results in increasing heat fluxes. To improve the efficiency and reliability of electronic systems, heat needs to be dissipated efficiently[1]. In terms of material selection, polymers offer several advantages over metals such as low cost, corrosion resistance, easy of moldability, and lower weight. High thermal conductivity polymer materials can improve thermal management in a wide range of applications, such as - water desalination[2], automotive control units[3], batteries[4], solar panels[5], supercapacitors[6], electronic packaging[7], and electronic cooling[8]. A key approach to enhance thermal conductivity of polymers is addition of high thermal conductivity fillers such as graphene ($k$ >2000 Wm$^{-1}$K$^{-1}$ [9, 10]). Different approaches have been used to enhance composite $k$ value through graphene, such as synergistic effect with multiple fillers[11] and alignment of graphene[12, 13]. The success of these approaches is, however, limited by the large interface thermal resistance between graphene and polymer in the range of $10^{-8}$ to $10^{-7}$ m$^2$ KW$^{-1}$ [13, 14] due to mismatch of phonons (lattice vibrations) between the two. To decrease thermal interface resistance, graphene is chemically functionalized by groups that are compatible with the surrounding polymer[15, 16]. Covalent functionalization[6] and non-covalent functionalization[17] of graphene can lead to higher interfacial thermal conductance. Two orders of magnitude increase in interface thermal conductance[16] and 156% enhancement[18] in composite thermal conductivity were achieved through grafting of polymer chains on graphene. Recent work demonstrated that multilayer graphene is more efficient at enhancing thermal conductivity than single layer graphene[19]. For such multilayer graphene, it is critically important to understand the optimal location of functional groups (such as edge or basal plane of graphene, Figures. 1a-d) which can lead to highest composite thermal conductivity. In this work we demonstrate that functionalization on the edges (Figures. 1b and c) can lead to significantly higher effective polymer thermal conductivity compared to functionalization on the basal plane (Figures. 1a and d). The functionalization scheme used in this work is oxidation of graphene since oxygen groups on graphene can interact with oxygen groups in polyetherimide through hydrogen bonding (Figure. 1e). Edge oxidation is achieved in this work using a recently introduced chemical pathway, by oxidizing graphene in presence of sodium chlorate and hydrogen peroxide to introduce an excess of carboxyl moieties[20]. Oxidation of the basal plane is achieved through Hummers method[21].

Several computational and experimental studies have reported enhancement of thermal conductivity through functionalization. Theoretical studies on polymer grafted graphene (performed by Wang *et al.*) showed two-fold lower interfacial thermal resistance through attachment of functional groups[22]. Using molecular dynamics (MD) simulations, Konatham *et al.*[23]

demonstrated almost 50% reduction in the interfacial thermal resistance between functionalized graphene and octane. Lin et al.[16] showed using MD simulations, a 22% increase in interface thermal conductance through use of alkyl-pyrene functional group. Ganguli et al.[24] showed that silane-functionalized graphene improved thermal conductivity of graphene/epoxy composite by 50% compared to pristine graphene composite for 8 weight% graphene content. Pyrene-end poly(glycidyl methacrylate) functionalized graphene/epoxy composite was found to yield a ~184% enhancement in thermal conductivity over pure epoxy[25]. Comparison between edge and basal plane functionalization has also been studied for several applications. Yang et al.[26] compared the role of edge and basal plane functionalization in modifying interfacial tension between graphene and liquids. Mungse et al.[27] studied the lubrication potential of basal plane alkylated graphene nanosheets. In another study, the effect of interconnected 3D network structure of edge or basal plane modified graphene oxide on electrical conductivity[28] was examined. Xiang et al.[29] compared edge versus basal plane functionalization of graphene for energy conversion and energy storage applications. There is, however, a lack of detailed understanding of the relative effectiveness of edge versus basal plane functionalization in enhancing thermal conductivity of polymer-graphene nanocomposites.

The advantage of edge-oxidation can be understood by observing that oxidation of the basal plane through Hummers method[21] distorts the planar $sp^2$-$sp^2$ carbon network of the basal plane by requiring a transition to tetrahedral $sp^3$ hybridization needed to accommodate the extra bond for oxidation (Figure 1a). These sites of distortion act as phonon scatterers, dramatically lowering the intrinsic thermal conductivity of graphene. In general, a subsequent reduction reaction is required to partially restore the $sp^2$ carbon-carbon network, before graphene can be used to enhance thermal conductivity of composites. Defective structure of basal plane oxidized graphene has been studied through X-ray diffraction (XRD), scanning tunneling microscopy (STM) and Raman scattering[30,31]. The oxidation related defects are found to be mainly due to the presence of hydroxyl (-OH) and epoxy (C-O-C) functional groups on the basal plane of graphene[30]. Bagri *et* al. also studied the defective state of GO and indicated the presence of hydroxyl and epoxy groups on the basal plane as the main defect sites[32]. We later show the presence of these functional groups in basal plane oxidized graphene through Fourier Transform Infrared (FTIR) spectroscopy, thermogravimetric analysis (TGA) and X-Ray photoelectron spectroscopy (XPS) in sections 3.5, 3.6 and 3.7, respectively. Reduction in thermal conductivity through such functionalization of the basal plane has been well studied computationally. While the intrinsic thermal conductivity of graphene is ~ 2000 $Wm^{-1}K^{-1}$, Fugallo *et al.* used first-principles computations to report that this value dropped by a factor of 2 in hydrogenated graphene and by one order of magnitude in fluorogenated graphene [33]. Mu *et al.* [34] used molecular dynamics simulations to predict dramatic decrease in thermal conductivity of basal plane oxidized graphene. An oxygen coverage of mere 5% was shown to reduce the *k* value of basal-plane oxidized graphene dramatically by 90% and a coverage of 20% lowered it further to ~8.8 $Wm^{-1}K^{-1}$, even lower than the amorphous limit. Shenogin *et al*.[35] similarly used MD simulations to show a decrease in *k* of carbon nanomaterials through functionalization. The much lower thermal conductivity of basal plane functionalized graphene limits its effectiveness in enhancing thermal conductivity of the polymer composite and can even result in a decrease in composite thermal conductivity relative to the case of pristine graphene. Reduction reaction with different reactants[36] or simultaneous amination[37] are used to partially restore the thermal conductivity of graphene. 15.8 wt% thermally reduced-RGO/epoxy composites (prepared using modified Hummers method followed by two step reduction reaction) yielded a *k* value of 1.27 $Wm^{-1}K^{-1}$ (relative to 0.275 $Wm^{-1}K^{-1}$ for pure epoxy)[38]. Amine functionalized GO/PEI composite reached thermal conductivity values up to 0.33 Wm$^{-1}$K$^{-1}$ at 3wt% phenyl aminated graphene-oxide filler concentration[39] (showing an enhancement of 43% relative to pure PEI thermal conductivity of 0.23 Wm$^{-1}$K$^{-1}$). Bernal *et al.* prepared graphene nano paper from phenol functionalized graphene followed by linking this phenol functionalized graphene with dianiline and directly measured a *k* value of 0.672 Wm$^{-1}$K$^{-1}$, showing 190% enhancement with respect to pristine GnP nanopaper[40].

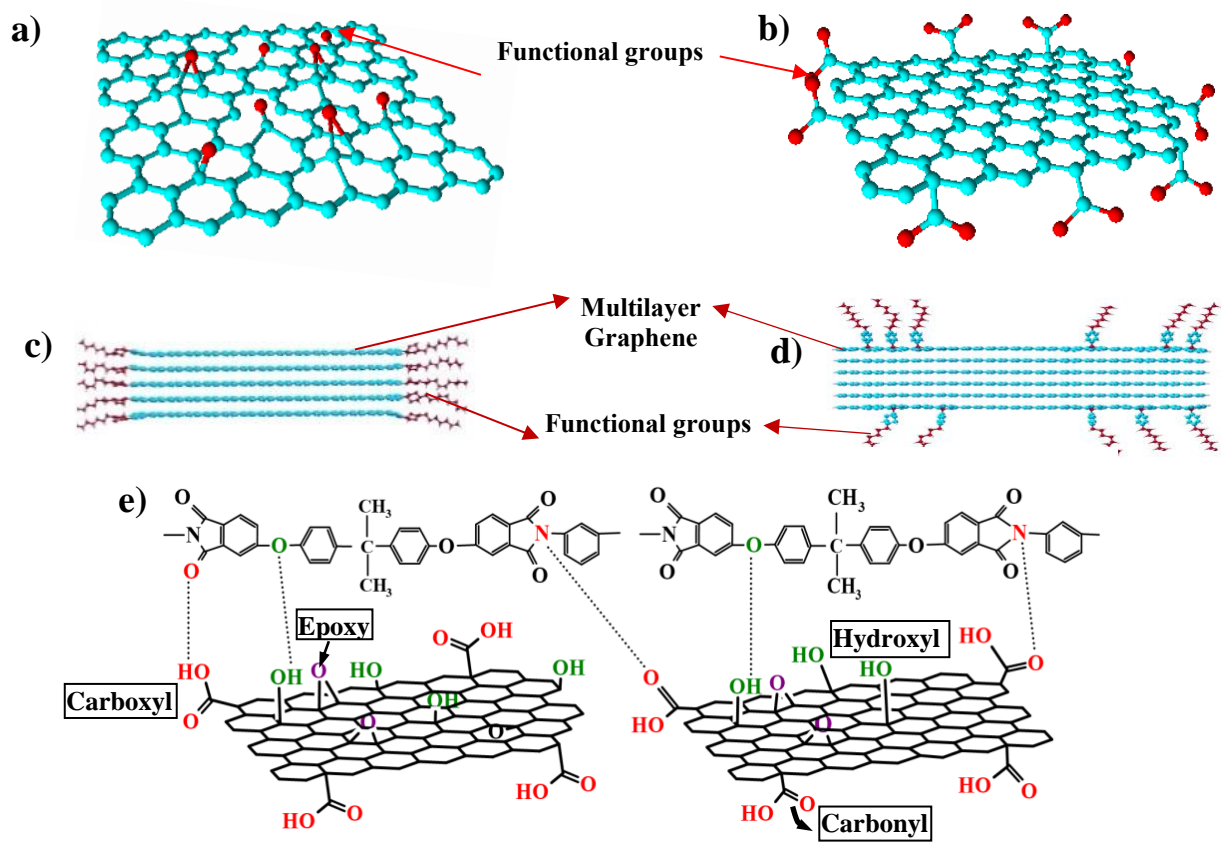

**Figure 1.** Atomic structure of a) functionalized single graphene sheet with basal-plane functional groups, b) functionalized single graphene sheet with edge functional groups. Atomistic structures for c) multilayer edge functionalized graphene (EFGnP) and (d) multilayer basal-plane functionalized graphene (BFGnP), e) interactions between graphene-oxide and polyetherimide (PEI).

Edge oxidation of graphene offers the advantage of leaving the basal plane intact (Figure 1b), preserving its ultra-high thermal conductivity (~ 2000 Wm$^{-1}$K$^{-1}$)[14]. Using MD simulations, the thermal conductivity of edge-functionalized graphene has been shown to be within 90% of the pristine graphene value[41]. Simultaneously, hydrogen bonding interaction between oxygen groups of graphene (on edge) and polyetherimide enhances interfacial thermal conductance between graphene and polyetherimide (PEI). The preserved high thermal conductivity of graphene coupled with superior interfacial thermal conductance with polymer can lead to superior enhancement in thermal conductivity of polymer composite through edge oxidation, compared to the case of basal plane oxidized graphene.

In our recent computational work[41], we also demonstrated other advantages of edge functionalization for enhancement of composite thermal conductivity. We showed that edge-bonding couples polymer to the high in-plane ($k_{in}$ > 2000 Wm$^{-1}$K$^{-1}$)[9] thermal conduction pathway of all graphene sheets within a nanoplatelet (Figure 1c) thus establishing a very efficient thermal conduction path through the composite. Basal plane bonding, on the other hand, primarily couples functional groups only to the outermost surface layers of the nanoplatelet (Figure 1d). The weak van der Waals coupling of outer layers with inner layers renders the inner layers of the nanoplatelet to be less efficient in conducting heat due to the poor through-thickness thermal conductivity of graphite (~10 Wm$^{-1}$K$^{-1}$)[13, 42] (Figure 1d). We further demonstrated that edge bonding leads to weak damping of vibrations within the graphene sheets, while for basal plane bonding, the coupling of the entire basal plane with embedding polymer was found to strongly dampen vibrations, further promoting the advantage of edge bonding. Finally, we demonstrated using *ab initio* atomistic Green's function computations[43], that interfacial thermal conductance at an individual junction between a functional group and graphene was higher, when the functional group was located on edge, as opposed to on the basal plane of graphene. This higher interfacial conductance is explained in terms of higher transmission to in-plane phonons of graphene for edge bonding case.

Above computational results provide a comprehensive understanding of the advantage of edge bonding in enabling superior thermal conductivity enhancement of composite. In this work, we demonstrate this advantage of edge bonding experimentally. Selective edge oxidation of graphene is obtained in this work by using the scheme outlined by Miao et al.[20] involving oxidizing graphene in presence of sodium chlorate and hydrogen peroxide in sulfuric acid. Miao et al.[20] showed that such a scheme leads to an excess of carboxyl groups on the edge of oxidized graphene. We confirm the presence of an excess of carboxyl groups through above oxidation scheme using FTIR and XPS analysis in sections 3.5 and 3.7, respectively. Carboxyl groups are known to preferentially form on the edge of graphene, yielding edge oxidized graphene. To experimentally show the edge localization of carboxyl groups, Yuge et al.[44], stained carboxyl groups using Pt-amine complex and found Pt-amine clusters to mainly exist at edges of graphene sheets. Computations based on density-functional theory also showed through geometric arguments that carboxyl groups are more likely to form on graphene edges[45]. This work is the first to report enhancement in thermal conductivity through the use of chemical edge oxidation pathway discussed above. Basal plane oxidation was achieved by using Hummers method[21] by oxidizing graphene in presence of sodium nitrate and potassium permanganate.

The preferential edge functionalization through the Miao's scheme[20] is confirmed in this work through location dependent 2-dimensional (2D) Raman mapping of nanoplatelets. Functionalized graphene is further characterized using X-ray diffraction (XRD), X-ray photoelectron spectroscopy (XPS), thermogravimetric analysis (TGA), Fourier transform infrared (FTIR) spectroscopy, and scanning and transmission electron microscopy (SEM and TEM) to examine the physical and chemical differences between the pristine, edge and basal plane oxidized graphene.

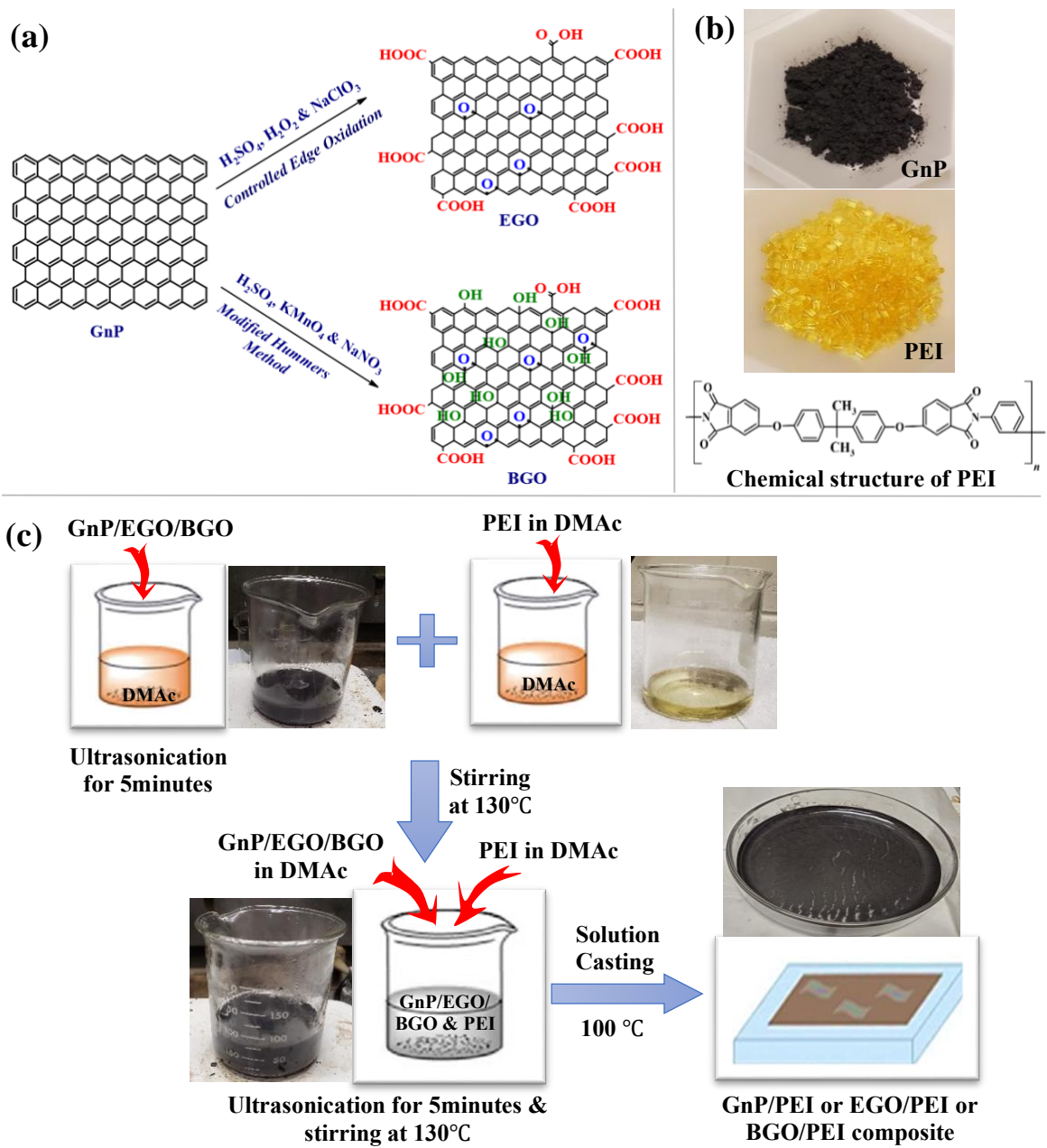

**Figure 2.** a) Schematic structure of GnP, EGO, and BGO (carboxyl & carbonyl group: -COOH/-C=O; epoxy: C-O-C; hydroxyl group: -OH), b) images of GnP nanopowder, PEI pellets and chemical structure of PEI; c) scheme for polymer-graphene composite preparation; DMAc-Dimethylacetamide

## 2. EXPERIMENTAL SECTION

**2.1 Materials.** Graphene nanoplatelets used in this work have an average thickness of ~60 nm and a lateral size of ~7 μm. The graphene nanopowder was purchased from Graphene Supermarket[46]. Potassium permanganate ($KMnO_4$, 99%), sulfuric acid ($H_2SO_4$, 95–98%), hydrogen peroxide ($H_2O_2$, 30%), sodium chlorate ($NaClO_3$, 99%), sodium nitrate ($NaNO_3$), N, N-dimethylacetamide (DMAc), hydrochloric acid (HCl, 35.0 -37.0%), and polyetherimide (PEI pellets, melt index 18 g/10 min) were purchased from Sigma Aldrich[47].

**2.2 Synthesis of edge functionalized graphene oxide (EGO).** The synthesis of edge oxidized graphene was performed with a controlled oxidation reaction using $NaClO_3$, $H_2SO_4$ and $H_2O_2$ (Fig. 2a) according to the approach outlined by Miao et al.[20]. 80 mL of $H_2SO_4$ was cooled down to 0°C temperature using an ice bath and 1gm of graphene-nanopowder was dispersed into the $H_2SO_4$ using bath sonication for 15 min. After dispersing graphene into $H_2SO_4$, approximately 6 mL of 30% $H_2O_2$ solution was added to the mixture and stirred at 0°C for few minutes. Then 4 gm of $NaClO_3$ was added very slowly and carefully into this mixture for 2 h. This mixture was kept stirring at room temperature for 8 hrs. The reaction mixture was then poured into 500 ml of cold DI water. The mixture was centrifuged at 4000 rpm to separate the particles from acidic solution. The separated particles were then washed twice with 800 ml of HCl aqueous solution (HCl:DI water = 1:9), followed by repeated filtration with ethanol, acetone and DI water until the pH reached neutral condition. The product was then kept in vacuum oven at 60 °C for 24 hrs. These oxidized particles are edge functionalized graphene oxide, denoted as EGO.

**2.3 Synthesis of basal-plane functionalized graphene oxide (BGO).** Hummers method[21] was used to prepare basal-plane functionalized graphene oxide (Fig. 2a). For this synthesis, 1 g graphene nanopowder was added into a mixture of 46 mL of sulfuric acid and 1 g of $NaNO_3$. This reaction was stirred for 4 h at 0°C in ice-cold bath to get a homogeneous dispersion. 6g KMnO$_4$ was added very slowly into this mixture. To maintain the temperature at 0 °C, the addition of KMnO$_4$ was done for 1 h. Then the reaction mixture is kept stirring at 35 °C. After 6 h, the reaction mixture was added to 92 mL of DI water at 95 °C and stirred for 15 min. In the last step, this mixture was mixed with 20% H$_2$O$_2$ aqueous solution. The final product was washed several times with HCl aqueous solution, ethanol, acetone and DI water repeatedly to remove ions and impurities. The separated particles were dried in a vacuum oven at 60 °C for 24 h to obtain the basal plane functionalized graphene oxide, named as BGO.

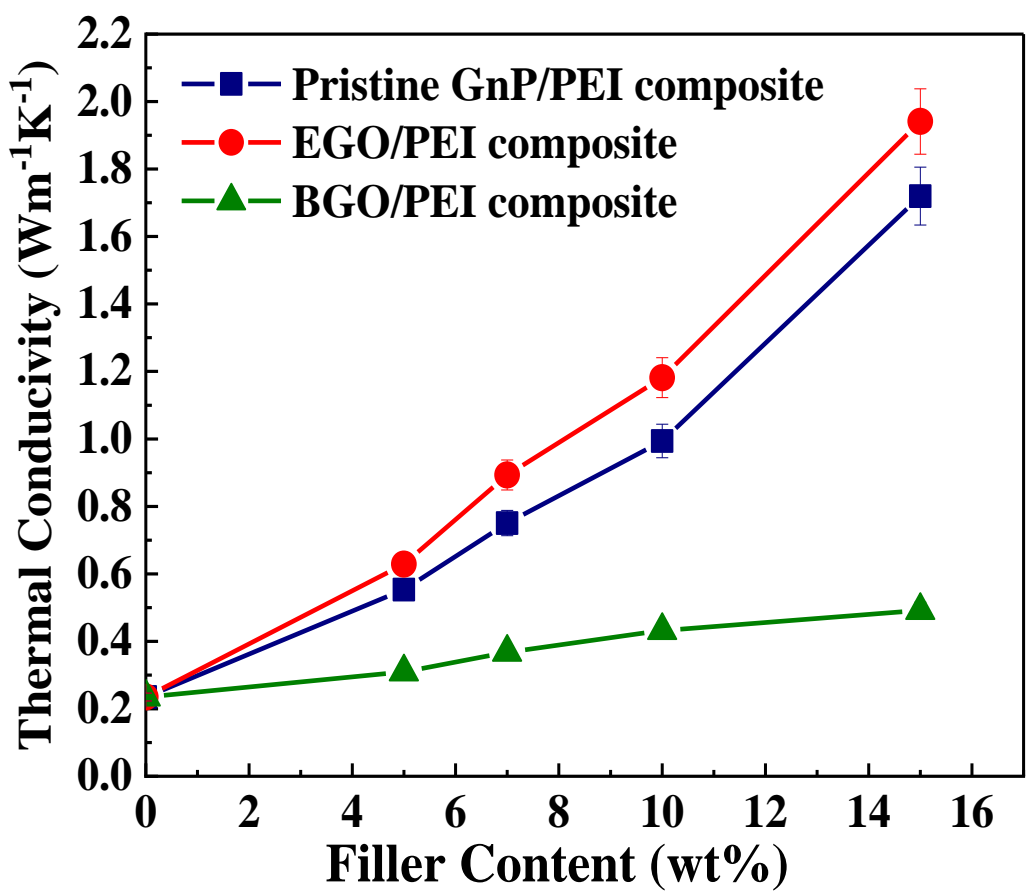

**Figure 3.** Through -thickness thermal conductivity value with 5, 7, 10, and 15 wt% filler content of GnP/PEI, EGO/PEI, BGO/PEI composites.

**2.4 Preparation of pristine GnP/PEI, EGO/PEI, and BGO/PEI composites.** A solution casting method was used to prepare the composites and the through-thickness thermal conductivity values of these samples were then measured and compared. Pristine 60 nm GnPs were dispersed into 20 mL DMAc for 30 min using a probe sonicator (Fig. 2c). Separately PEI pellets were dissolved into 50 mL DMAc at 130 °C for 1 h. Dispersed graphene solution was blended into dispersed polymer solution for 3 h at 130 °C. To disperse the graphene into polymer properly, graphene and polymer solutions were mixed together and ultrasonicated for 5 mins. The homogenized solution was cast into a petri dish and held at 100 °C for 24 h producing a graphene-PEI composite film. 7 wt%, 10 wt%, and 15 wt% GnP-PEI, EGO-PEI, and BGO-PEI composite films were prepared using the same procedure.

## 3. RESULTS AND DISCUSSION

**3.1 Thermal Conductivity Data.** Figure 3 presents the comparison of $k$ values of GnP/PEI, EGO/PEI and BGO/PEI composites for different GnP filler loadings. Thermal conductivity ($k$) value of pure polyetherimide is measured to be 0.23 Wm$^{-1}$K$^{-1}$ (in good agreement with literature value[48]), while thermal conductivity of pristine graphene-polyetherimide composite (shown by blue curve in Figure 3) is measured to be 1.72 Wm$^{-1}$K$^{-1}$ for 15 weight% filler content. Figure 3 further shows that the highest thermal conductivity values are achieved through the use of edge-oxidized graphene (EGO), clearly demonstrating the advantage of edge-oxidation. Enhancements in $k$ value of 18% and 13% are achieved for EGO/PEI composite relative to pristine GnP/PEI composite, resulting in high $k$ values of 1.14 Wm$^{-1}$K$^{-1}$ and 1.94 Wm$^{-1}$K$^{-1}$ for EGO/PEI composite for 10 wt% and 15 wt% graphene, respectively. The $k$ value of 1.94 Wm$^{-1}$K$^{-1}$ achieved for 15 wt% EGO/PEI composite, represents a large ~725% enhancement compared to pure polyetherimide.

On the other hand, $k$ value of BGO/PEI composite is found to be diminished relative to that of pristine GnP/PEI composite. At 15 weight% composition, the $k$ value of BGO/PEI composite (0.48 Wm$^{-1}$K$^{-1}$) is lower by almost 72% and 75% relative to pristine GnP/PEI and EGO/PEI composites, respectively. The BGO/PEI $k$ is still found to be enhanced with respect to pure PEI by ~104% at 15 weight% composition, but by a significantly smaller margin of 725% achieved for EGO/PEI composite.

The lower $k$ of BGO/PEI composite is due to the defects introduced by Hummers method on the basal plane of GnP[49] (as discussed before), which dramatically lower the intrinsic thermal conductivity of graphene. We later provide evidence (through $I_D/I_G$ ratio obtained via Raman analysis in section 3.3) of the presence of much larger number of functional groups on the basal plane of graphene for BGO compared to pristine GnP and EGO. XRD analysis (presented in section 3.4) further shows that these oxygen functional groups lead to an increase in interlayer spacing of BGO to 0.717 nm (relative to values of 0.336 nm for pristine GnP) indicating that the oxygen groups intercalate the spacing between the graphene layers within the nanoplatelet for BGO and attach to the basal planes of all graphene sheets constituting the nanoplatelet. Previous computations have shown that such basal plane oxidized graphene has a dramatically diminished $k$ compared to pristine graphene (by up to 90% lower for 5% coverage)[34]. The significantly lower $k$ of BGO compared to pristine graphene causes the $k$ of BGO/PEI composite to become even lower than that of pristine GnP/PEI composite.

For edge bonding (EGO), however, we show through Raman analysis (in section 3.3), a much smaller presence of functional groups on the basal plane of graphene. XRD analysis (discussed in section 3.4) further shows that interlayer spacing of EGO (0.337 nm) is very similar to that of pristine graphene (0.336 nm), providing more evidence of the presence of oxygen groups on edges of graphene for EGO. XPS analysis (section 3.7) further shows a stronger intensity of C-C/C=C peak in EGO compared to BGO. These analysis point to superior structural integrity of the basal plane of graphene for EGO relative to BGO. Computational analysis reveals that $k$ of such structurally preserved EGO can be close to the high intrinsic $k$ of pristine graphene. Simultaneously, hydrogen bonding between oxygen groups of EGO and PEI enhances interfacial thermal conductance between the two. Significantly superior *k* of EGO (and close to the value of pristine graphene) compared to BGO, combined with improved interfacial thermal conductance between EGO and PEI (relative to between pristine GnP and PEI) causes *k* of EGO/PEI to exceed that of both pristine GnP/PEI and BGO/PEI composites (Figure 3). These results can enable new avenues to achieve higher thermal conductivity polymer/graphene nanocomposites. Below we provide understanding of dispersion of graphene in the prepared composite. We further provide understanding of differences in oxidation in EGO and BGO through detailed characterization.

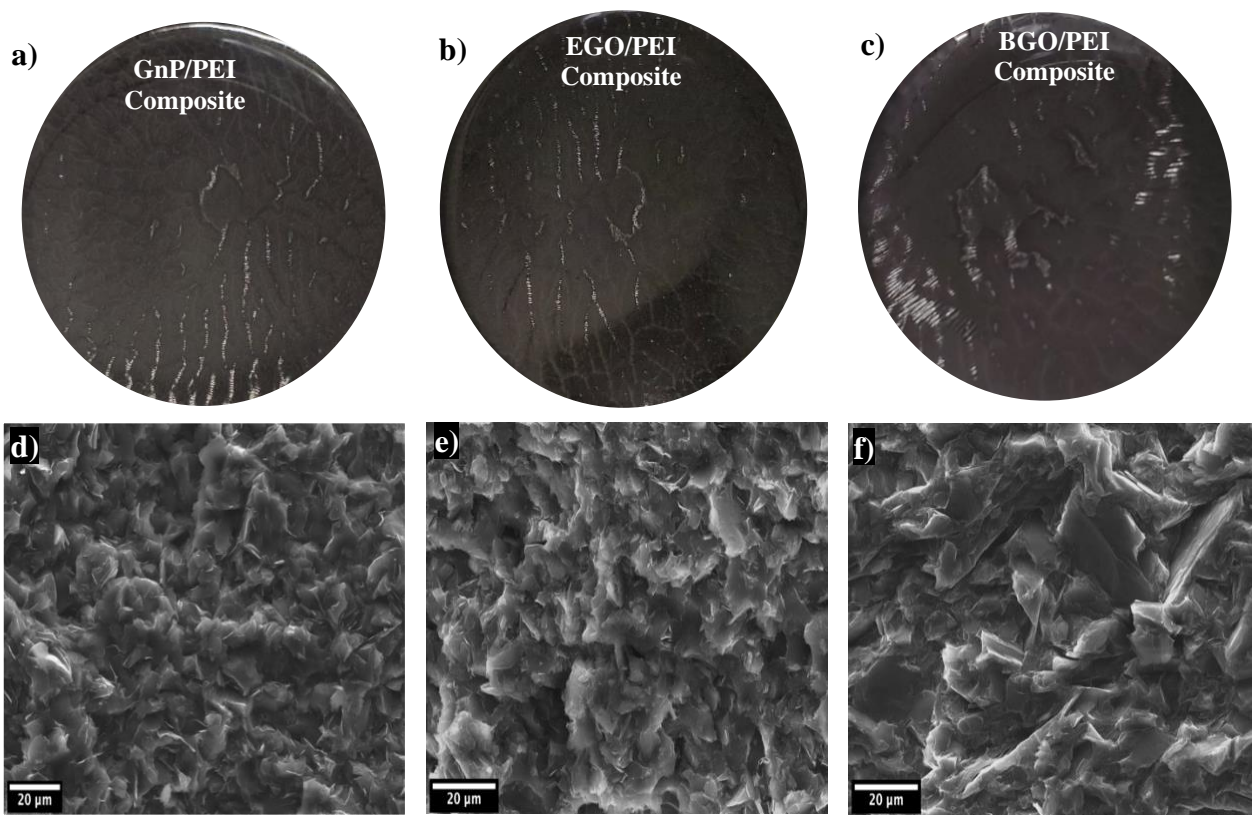

**Figure 4.** Images of composite films for a) pristine GnP/PEI, b) EGO/PEI, c) BGO/PEI & cross-sectional FE-ESEM images of 15wt% composites for d) pristine GnP/PEI, e) EGO/PEI and f) BGO/PEI composites.

**3.2 Dispersion of Graphene nanoplatelets within the composite.** Uniform dispersion of graphene in the polymer matrix is critical for achieving high thermal conductivity. In this work, we used N, N-dimethylacetamide (DMAc) to both dissolve polyetherimide and disperse graphene nanoplatelets. DMac is a highly polar aprotic solvent and has a very strong dispersing power. This enables significantly lower aggregation compared to other solvents such as acetone or ethanol. Figures. 4(a-c) show that graphene nanoplatelets were uniformly dispersed in all three composites – pristine GnP/PEI, EGO/PEI and BGO/PEI. While Figures. 4a-c show dispersion of graphene in bulk samples, we also performed cross-sectional FE-SEM analysis to study dispersion of graphene on a microscopic scale. These images are shown in Figure 4d-f for 15wt% GnP/PEI, EGO/PEI and BGO/PEI composites. These images clearly indicate that pristine GnP, EGO and BGO fillers are all similarly well dispersed into polymer matrix even at a microscopic level. The similar uniform dispersion of graphene for all three cases (pristine, edge-oxidized and basal-plane oxidized) indicates that the advantage of edge-bonding in enhancing *k* relative to pristine graphene, can be attributed to superior interfacial thermal interaction with the surrounding polyetherimide.

**3.3 Raman Characterization.** To provide evidence of selective edge oxidization of graphene, 2D Raman mapping of EGO and BGO (Figure 5a-c) was performed. Raman spectroscopy characterizes graphene's unique structure through two conventional peaks, one at ~1586 cm$^{-1}$ another one at ~1345 cm$^{-1}$, attributed to G band and D bands respectively[50]. The presence of G band is related to the stretching mode of defect-free sp$^2$ carbon through the first order E$_{2g}$ scattering mode and the D band is induced by the disordered structure of sp$^3$ hybridized carbon[51]. $I_D/I_G$ (the ratio of the intensity of D band and G band)[52] thus provides an understanding of the structural disorder of graphitic structure.

To examine and compare the $I_D/I_G$ ratio at the "edge" vs. "basal plane" of both edge (EGO) and basal plane (BGO) oxidized graphene, a DXR3 Raman microscope is used and graphene nanoplatelets (pristine graphene/edge oxidized graphene/basal plane oxidized graphene) are transferred on a glass slide on top of a piece of double-sided tape to ensure that they do not move/drift during the Raman mapping. For treated/untreated graphene, 1-2 graphene nanoplatelets are selected for Raman area mapping. Snapshot of pristine graphene nanoplatelet captured during the 2D Raman mapping using a 100x objective is shown in Figure S1a. The red rectangle (as shown in Figure S1a) is the area (10 μm x 4.2 μm) selected for Raman mapping, with each red dot representing a spot for Raman spectral collection (total 147 points, with a step size of 0.5μm). The spectra were collected from points on edge and basal plane area of graphene nanoplatelets separately to compare the $I_D/I_G$ ratio of edge and basal plane for pristine GnP, EGO and BGO. The process involved in 2D Raman mapping is presented in supporting information in Figure S1a-d.

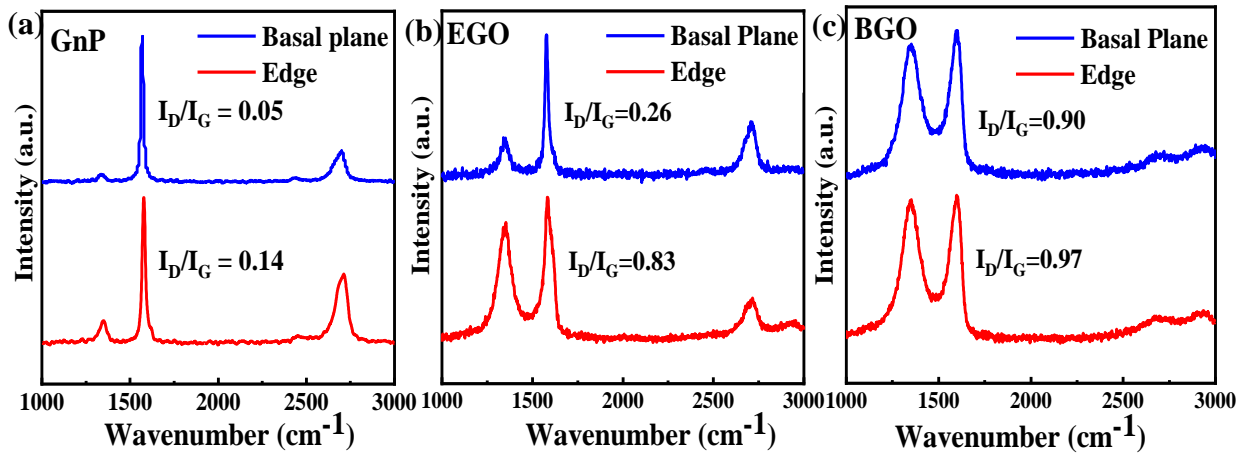

**Figure 5.** $I_D/I_G$ ratio of edge and basal plane area of (a) GnP, (b) EGO, and (c) BGO.

These $I_D/I_G$ ratios for the center area and edge are compared in Figures. 5 (a-c) for pristine GnP, edge-oxidized GO (EGO) and basal-plane oxidized GO (BGO). $I_D/I_G$ ratio of EGO increases more dramatically on edges (0.83) compared to $I_D/I_G$ ratio of pristine graphene (0.14 for edges), relative to the increase on basal plane (0.26 for EGO versus 0.05 for pristine case). This much larger increase of $I_D/I_G$ ratio on edges suggests selective edge oxidation of edges of EGO. However, for BGO, increase in $I_D/I_G$ ratio on basal plane is more significant (0.90 for BGO versus 0.05 for pristine case) suggesting basal plane oxidation for BGO.

The higher $I_D/I_G$ ratio on the center area of BGO (0.90) compared to EGO (0.25) indicates greater number of functional groups on center area of BGO relative to EGO; these functional groups distort the basal plane of graphene in BGO[26, 53], causing a reduction in its thermal conductivity which in turn decreases the overall thermal conductivity of BGO/polyetherimide composite (Figure 5c). Overall $I_D/I_G$ ratio (for the entire nanoplatelet) of EGO (0.83) is also significantly lower than BGO (1.02) indicating overall less structural damage in EGO than BGO (Figure 6a). More details of Raman analysis are presented in the supporting information (Table S1).

**3.4 Effect of functionalization on interlayer spacing through XRD analysis.** To further provide evidence of distortion of graphitic structure in BGO, we performed XRD analysis to obtain an understanding of interlayer spacing for the two different oxidation schemes. The interlayer spacing is measured from Bragg's law[54] by using,

$$n\lambda = 2d \sin \theta \quad (1)$$

where, λ is the X-ray wavelength (0.15404 nm), and θ is the diffraction angle (radians). XRD diffraction pattern of pristine GnP shows a sharp reflection peak (002) at 2θ = 26.5° [55] (Figure 6b), pointing to an interlayer spacing (d) of 0.336 nm (computed from Eq. 2) for pristine GnP. For edge-oxidized nanoplatelet (EGO), the diffraction peak is still sharp and intense at the same location (2θ = 26.4°) as pristine nanoplatelet, indicating a similar interlayer spacing of 0.337nm in EGO as pristine graphene. This indicates that the functionalization scheme based on Miao's approach[20] mostly accessed the edge of graphene without penetrating the interlayer spacing. XRD pattern for EGO also exhibits a very weak peak at 2θ = 13.5° [56], this corresponds to the introduction of oxygen groups at the edge of graphene. This localization of functional groups on graphene edges for EGO, leaves the basal plane mostly intact, preserving the high thermal conductivity of graphene.

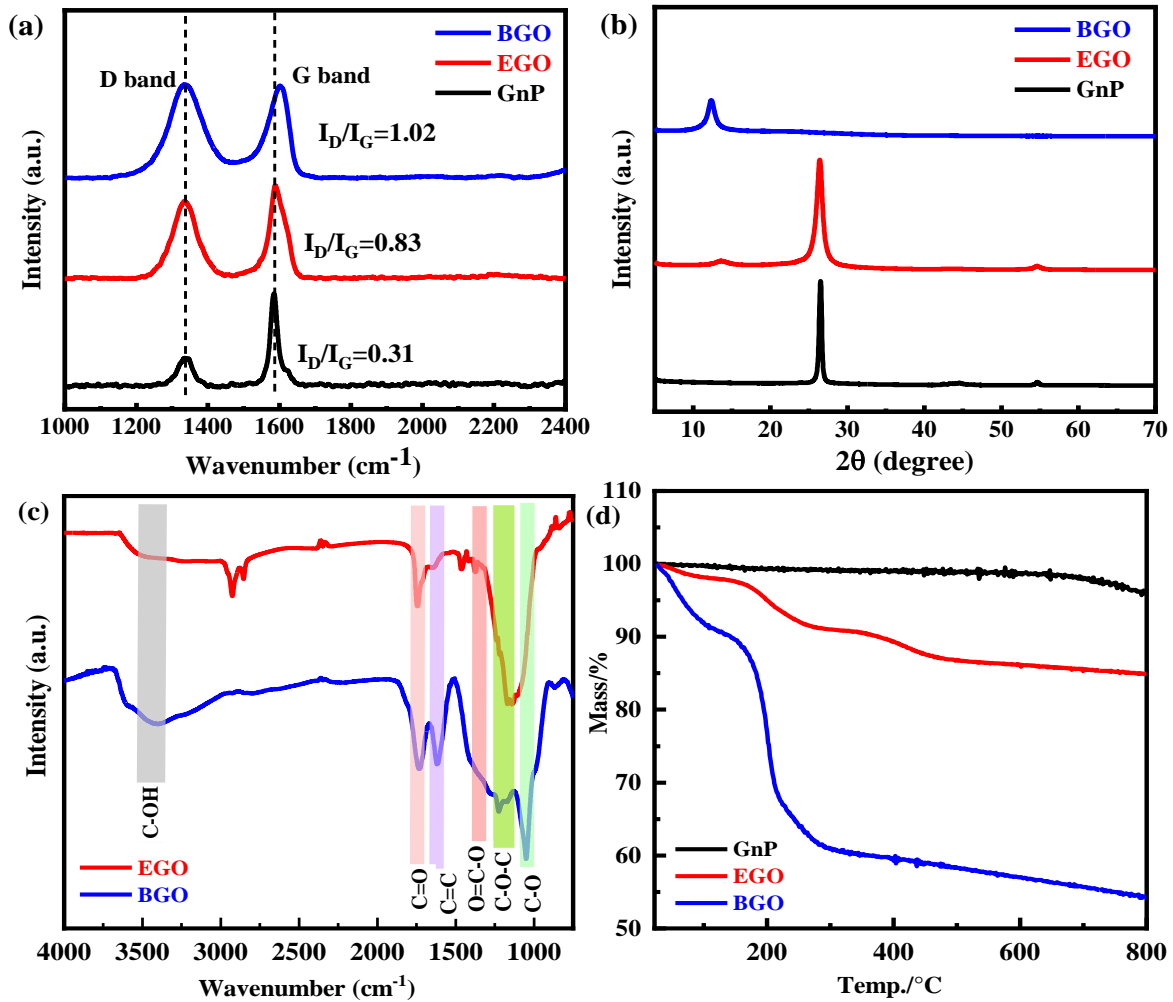

**Figure 6.** (a) Raman spectra, (b) Full XRD spectra (c) FTIR, and (d) TGA spectra of GnP, EGO and BGO.

For graphene oxidized by Hummers method (BGO), it is observed that the reflection peak 2θ = 26.5° completely disappears and shifts to 12.33° indicating higher interlayer spacing of 0.717 nm; this higher spacing is due to the intercalation of oxygen functional groups in between graphene layers in the nanoplatelets, suggesting oxidation of basal plane of all graphene sheets within the nanoplatelets for BGO, resulting in a disordered graphitic structure[57]. Such basal plane oxidized graphene has a dramatically lower thermal conductivity relative to pristine graphene[34].

**3.5 Chemical group analysis through FTIR analysis.** FTIR analysis was carried out in ATR (attenuated total reflection) mode to observe the difference in functional groups for EGO and BGO. Oxidation of graphene leads to the presence of epoxy(C-O-C) and hydroxyl (C-OH) functional groups on the basal plane and carbonyl (C=O) and carboxyl (-COOH) functional groups on the edge according to Lerf–Klinowski model[58] (see Figure 1e). Figure 6(c) shows that EGO exhibits an intense peak at ~1740 cm$^{-1}$ [57] and a peak at ~1380 cm$^{-1}$ [59], corresponding to stretching vibrations of (C=O) and (C-O); these confirm the presence of carboxyl (-COOH) groups. BGO shows sharp and intense peak for epoxy group (C-O-C) at ~1231 cm$^{-1}$, [39] and a comparatively broader peak for hydroxyl (-OH) group at ~3400 cm$^{-1}$ [60] quite dissimilar from EGO. Outlined FTIR analysis reveals the difference in functional groups present in EGO and BGO. Carboxyl group formation is strongly noticeable in EGO confirming the introduction of an excess of carboxyl groups through Miao's oxidation scheme (involving use of sodium chlorate and hydrogen peroxide)[20].

**3.6 Thermogravimetric analysis.** Functionalization behavior is also observed with TGA (thermogravimetric) analysis in Figure 6d. Pristine GnP exhibits minor weight change with temperature indicating presence of a very small quantity of functional groups. Lower total weight loss for EGO (15wt%) compared to BGO (~45%) represents smaller quantities of functional groups present in EGO compared to BGO. Weight loss in EGO is found to occur very slowly up to 800 °C, indicating removal of highly stable oxygen groups which are usually edge functional groups[61]. Quick degradation in BGO from 110-300°C temperature range is not observed in EGO, which implies that BGO possesses higher amount of less stable epoxy and hydroxyl groups in comparison to EGO[28].

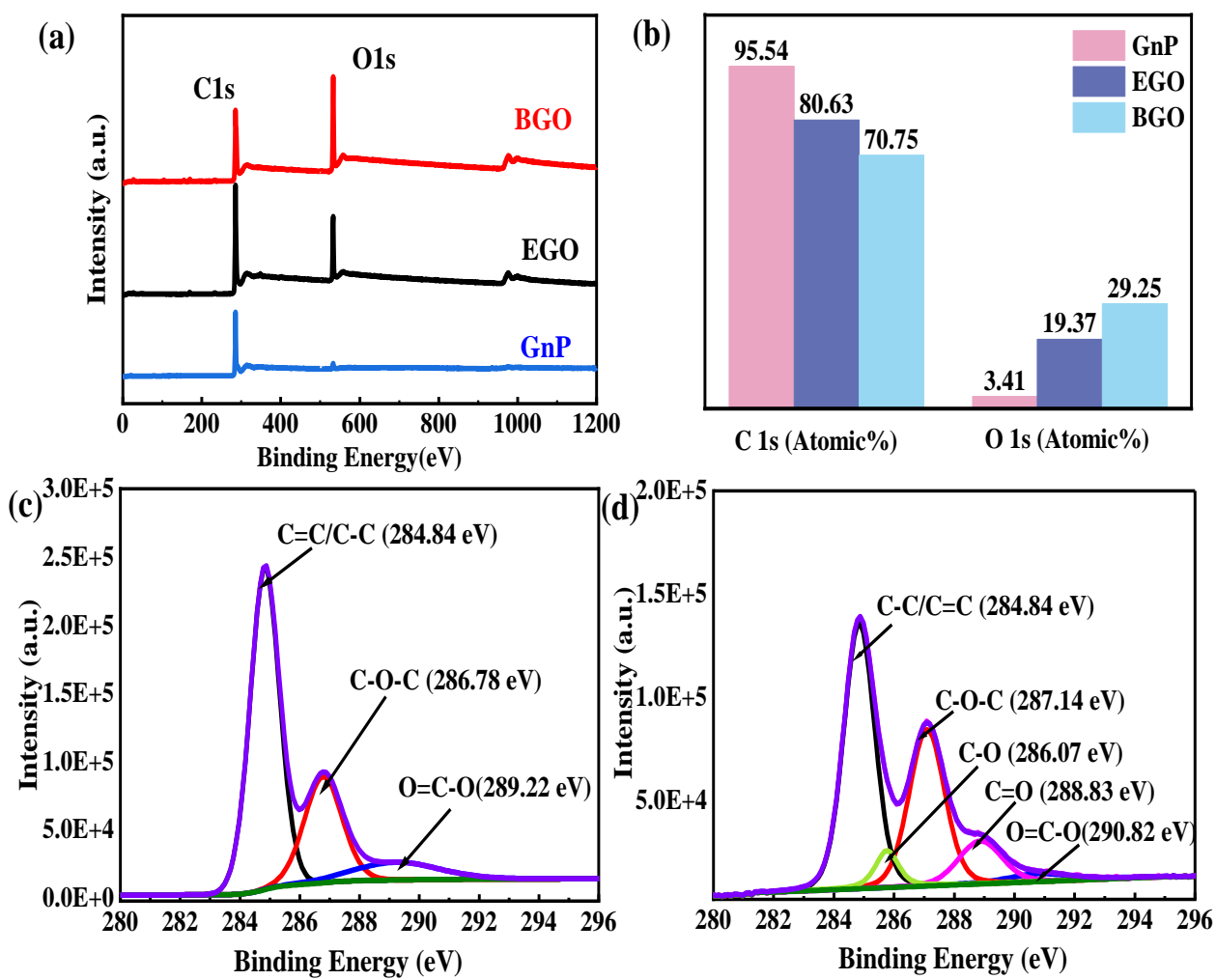

**Figure 7.** XPS data showing the (a) survey spectra, (b) atomic percentage of C1s and O1s for GnP, EGO, and BGO, and the high resolution C1s spectra for (c) EGO and (d) BGO.

**3.7 XPS analysis.** XPS analysis (seen in Figure 7) is further used to investigate the differences in oxidation in EGO and BGO samples. Figure 7a shows presence of C1s peak around 285 eV for pristine GnP, and strong peaks of O1s around 533eV for EGO and BGO[39]. The intensity of oxygen peak in BGO is higher than EGO suggesting a higher atomic percentage of oxygen functional groups in BGO compared to EGO (Figure 7b). The O/C ratios for EGO and BGO are 0.24 and 0.41 respectively.

**Table 1.** Abundance of functional groups of EGO and BGO in atomic percentage

|     | C-C/C=C | C-OH | C-O-C | C=O | O=C-OH |
| --- | --- | --- | --- | --- | --- |
| EGO | 62.99 | - | 26.87 | - | 10.14 |
| BGO | 46.28 | 8.49 | 33.48 | 10.25 | 1.51 |

To understand the differences in functional groups in EGO and BGO, the C1s high resolution XPS spectra was further analyzed. The deconvoluted spectra of BGO (shown in Figure 7d) reveals the presence of non-oxygenated peak of C-C/C=C at 284.84 eV, C-O-C (epoxy) at 287.14 eV, C-OH (hydroxyl) at 286.07 eV, C=O (carbonyl) at 288.83 eV, and O=C-OH (carboxyl) at 290.82 eV[62, 63]. Table 1 reveals that atomic percentage of epoxy and hydroxyl functional groups (as shown in Table 1) in BGO is higher than in EGO. These functional groups penetrate the graphitic structure modifying the basal plane of BGO, as indicated by the weaker intensity of C-C/C=C in BGO relative to EGO. Such modification of basal plane in BGO lowers the thermal conductivity of BGO relative to pristine GnP.

Contrary to the case of BGO, Figure 7c shows that EGO exhibits a significant presence of carboxyl groups which are known to be predominantly located at the edges of graphene[44, 45]. This combined with a presence of smaller quantity of epoxy groups (C-O-C peak at 286.78 eV[63, 64]) and the absence of hydroxyl groups (C-OH) in EGO, preserves the structural integrity of the basal plane of graphene in EGO. The stronger intensity of C-C/C=C peak in EGO (Table 1) further confirms the superior integrity of graphene in EGO through the controlled edge oxidation scheme[20] used in this work. The minimal distortion of basal plane of graphene in EGO preserves its high intrinsic thermal conductivity.

## 4. CONCLUSION

In summary, we have demonstrated superiority of edge-oxidation of graphene (EGO) in enhancing thermal conductivity of graphene-nanoplatelet (GnP)/polyetherimide (PEI), relative to oxidation of the basal plane (BGO). To achieve edge-oxidation, graphene was reacted with sodium chlorate and hydrogen peroxide in sulfuric acid, introducing an excess of carboxyl groups on graphene edges. Basal plane oxidation was achieved using Hummers method. Combined effects of the preservation of high intrinsic thermal conductivity of graphene and an enhancement in interfacial thermal conductance between polymer and graphene through edge oxidation, caused the thermal conductivity of EGO/PEI composite to exceed that of pristine GnP/PEI composite by almost 18% at 10 weight% graphene composition. Oxidation of the basal plane, however, dramatically lowers the intrinsically thermal conductivity of graphene, causing the $k$ of BGO/PEI composite to be diminished by 57% and 63% relative to pristine GnP/PEI and EGO/PEI composites, respectively at the same composition. Relative to pure PEI, edge-oxidized graphene was found to enhance composite $k$ by 725% at 15 weight% composition, while the corresponding enhancement through basal plane oxidation was only ~75%. Raman spectroscopy, XPS analysis, XRD and FTIR provided evidence for the preferential edge oxidation in EGO through a presence of an excess of carboxyl (-COOH) functional groups on graphene edges. These results open up new avenues to achieve higher thermal conductivity polymer-graphene nanocomposites, with important implications for a wide range of thermal management technologies.

**Measurement of through-thickness thermal conductivity.** Thermal conductivity measurements were performed though the laser flash technique. A Netzsch LFA 467 Hyperflash laser was used to measure the through-thickness thermal diffusivity of the samples. A short, pulsed laser beam is impinged on one side of the sample and the temperature is measured on the opposing side of the sample as a function of time. The film samples were cut into a 12.5 mm diameter circles and coated with a thin layer of graphite paint to increase the emissivity of the samples. Measurements were performed at room temperature (23 °C) and a total of 6-8 shots per sample were taken for each sample. The thermal diffusivity was determined using the following equation:

$$\alpha = (0.1388\, d^2)/t_{1/2} \quad (2)$$

where, $\alpha$ is the thermal diffusivity, $t_{1/2}$ is the time to obtain half of the maximum temperature on the rear surface, and d denotes the sample thickness. The thermal conductivity was calculated based on the following equation:

$$k = \alpha \rho C_p \quad (3)$$

where $k$, $\rho$, and $C_p$ represent the thermal conductivity, density, and specific heat of the sample, respectively. In this work, density and specific heat of the sample are calculated from the rule of mixtures.

**Molecular Structure Characterization.** A DXR3 Raman microscope from Thermo Fisher Scientific was used to collect the Raman spectra (shown in Figure 5). The following parameters were used to collect the spectra - laser wavelength λL = 532 nm, laser power at the sample: 0.5 mW and microscope objective: 100x. A 25 mm confocal pinhole aperture and 10 seconds collection for each spectrum was used. The Raman spectra shown in Figure 6a have been obtained using a Horiba Jobin-Yvon labRam HR instrument. Data were collected over the range from 2400 to 1000 cm-1 using a laser wavelength λL = 633 nm, spectral resolution = 0.16 cm-1, and imaging resolution = 702 nm. An Olympus BX 41 microscope with 5x objective, 10 s exposure time for 15 accumulations, and 3 scans per sample were used to collect the spectra.

A Rigaku SmartLab 3kW (Rigaku Corporation, Japan) was used to produce the X-ray powder diffraction (XRD) patterns of GnP, EGO, and BGO at room temperature using Cu Kα radiation (λ = 1.5406 Å) with a scan range of 2-8° min-1 and step size of 0.02°. Bragg-Brentano configuration was used to collect the data at room temperature and the operating parameters were applied over 2θ = 5 to 70°.

GnP, EGO and BGO were analyzed by Thermo Scientific K-alpha X-ray Photoelectron Spectroscopy (XPS) where Al Kα gun source was used to excite the sample and measurement was carried out for acquisition time of ~48 s at 400 μm spot size. The passing energy of 50 eV was utilized to find the C, O peak in this analysis spectrum. The elemental analysis of C & O as well as the abundance of functional groups were investigated using the Avantage software. To determine the functional groups peak position, FWHM (Full width at half maximum) and atomic percentage, Avantage software was used to do curve fitting utilizing Gaussian and Lorentzian functions.

Fourier Transform Infrared spectroscopy (FT-IR) spectra were collected on the GnP, EGO, and BGO samples using a Paragon 1000 FT-IR Spectrometer (PerkinElmer, Inc) with a germanium crystal in attenuated total reflectance (ATR) mode. Data were measured over a wavenumber range of 4000 to 500 cm-1 and Omnic software was used for spectral analysis.

Thermogravimetric analysis (TGA) was performed using NETZSCH TG 209 F1 Libra to evaluate the thermal stability of the graphene samples and functional groups attached during oxidation process. Data were collected over the temperature range of 50−800°C. N2 gas atmosphere was used with a heating rate of 10 °C/min.

Field Emission Environmental Scanning Electron Microscopy (FE-ESEM) was used to characterize the composite structure. A ThermoFisher Quattro S FE-ESEM was operated at 20 kV to perform the analysis of the samples. Morphological characterization of pristine GnP, EGO and BGO was carried out by scanning electron microscopy (SEM) and high-resolution transmission electron microscopy (HR-TEM) (as presented in Figure S2 and S3 respectively). Zeiss NEON 40 EsB Crossbeam instrument and JEOL 2000-FX 200kV Transmission Electron Microscope with the camera of acquisition of DE (Direct Electron)-12 were used to perform the high-resolution SEM and TEM analysis respectively. This field-emitting scanning electron microscope was operated at an accelerating voltage of 5 kV. To prepare the samples for SEM imaging, DMAc solvent was used to coat the smooth silicon surface and 300 mesh lacy carbon copper grid was used for TEM.

ASSOCIATED CONTENT

**Supporting Information.** 2D Raman mapping procedure; Full Width at Half Maximum (FWHM) for D & G band of GnP, EGO and BGO; SEM & TEM images of GnP, EGO and BGO.

AUTHOR INFORMATION

**Corresponding Author**

**\*Jivtesh Garg**- School of Aerospace and Mechanical Engineering, University of Oklahoma, Norman, OK, USA 73019; Email: garg@ou.edu

**Authors**

**Fatema Tarannum**- School of Aerospace and Mechanical Engineering, University of Oklahoma, Norman, OK, USA 73019

**Rajmohan Muthaiah**- School of Aerospace and Mechanical Engineering, University of Oklahoma, Norman, OK, USA 73019

**Swapneel Danayat**- School of Aerospace and Mechanical Engineering, University of Oklahoma, Norman, OK, USA 73019

**Kayla Foley**- Department of Chemical Engineering, University of Arkansas, Fayetteville, AR, USA 72701

**Roshan S. Annam**- School of Aerospace and Mechanical Engineering, University of Oklahoma, Norman, OK, USA-73019

**Keisha B. Walters**- Department of Chemical Engineering, University of Arkansas, Fayetteville, AR, USA 72701

**Notes**

There are no conflicts of interest to declare.

ACKNOWLEDGMENT

RM, FT, SD and JG acknowledge support from National Science Foundation CAREER award under Award No. #1847129. We also thanks Mohammed Ibrahim, PhD, from Thermo Fisher Scientific for collecting the Raman spectra for the 2D composite mapping.

REFERENCES

(1) Bar-Cohen, A. Thermal management of on-chip hot spots and 3D chip stacks. In *2009 IEEE International Conference on Microwaves, Communications, Antennas and Electronics Systems*, 2009; IEEE: pp 1-8. Saadah, M. A. Thermal Management of Solar Cells. UC Riverside, 2013.

(2) Dreiser, C.; Bart, H.-J. Mineral scale control in polymer film heat exchangers. *Applied thermal engineering* **2014**, *65* (1-2), 524-529.

(3) Mallik, S.; Ekere, N.; Best, C.; Bhatti, R. Investigation of thermal management materials for automotive electronic control units. *Applied Thermal Engineering* **2011**, *31* (2-3), 355-362.

(4) Lee, J.-K.; Lee, Y.-J.; Chae, W.-S.; Sung, Y.-M. Enhanced ionic conductivity in PEO-LiClO4 hybrid electrolytes by structural modification. *Journal of electroceramics* **2006**, *17* (2), 941-944.

(5) Huynh, W. U.; Dittmer, J. J.; Alivisatos, A. P. Hybrid nanorod-polymer solar cells. *science* **2002**, *295* (5564), 2425-2427.

(6) Luo, B.; Liu, S.; Zhi, L. Chemical approaches toward graphene-based nanomaterials and their applications in energy-related areas. *Small* **2012**, *8* (5), 630-646.

(7) Lu, X.; Xu, G. Thermally conductive polymer composites for electronic packaging. *Journal of applied polymer science* **1997**, *65* (13), 2733-2738.

(8) Naghibi, S.; Kargar, F.; Wright, D.; Huang, C. Y. T.; Mohammadzadeh, A.; Barani, Z.; Salgado, R.; Balandin, A. A. Noncuring graphene thermal interface materials for advanced electronics. *Advanced Electronic Materials* **2020**, *6* (4), 1901303.

(9) Balandin, A. A. Phononics of graphene and related materials. *ACS nano* **2020**, *14* (5), 5170-5178.

(10) Colonna, S.; Battegazzore, D.; Eleuteri, M.; Arrigo, R.; Fina, A. Properties of Graphene-Related Materials Controlling the Thermal Conductivity of Their Polymer Nanocomposites. *Nanomaterials* **2020**, *10* (11), 2167.

(11) Barani, Z.; Mohammadzadeh, A.; Geremew, A.; Huang, C. Y.; Coleman, D.; Mangolini, L.; Kargar, F.; Balandin, A. A. Thermal properties of the binary-filler hybrid composites with graphene and copper nanoparticles. *Advanced Functional Materials* **2020**, *30* (8), 1904008. Cui, X.; Ding, P.; Zhuang, N.; Shi, L.; Song, N.; Tang, S. Thermal conductive and mechanical properties of polymeric composites based on solution-exfoliated boron nitride and graphene nanosheets: a morphology-promoted synergistic effect. *ACS applied materials & interfaces* **2015**, *7* (34), 19068-19075.

(12) Tarannum, F.; Muthaiah, R.; Annam, R. S.; Gu, T.; Garg, J. Effect of Alignment on Enhancement of Thermal Conductivity of Polyethylene–Graphene Nanocomposites and Comparison with Effective Medium Theory. *Nanomaterials* **2020**, *10* (7), 1291. Yan, H.; Tang, Y.; Long, W.; Li, Y. Enhanced thermal conductivity in polymer composites with aligned graphene nanosheets. *Journal of Materials Science* **2014**, *49* (15), 5256-5264. DOI: 10.1007/s10853-014-8198-z.

(13) Saeidijavash, M.; Garg, J.; Grady, B.; Smith, B.; Li, Z.; Young, R. J.; Tarannum, F.; Bekri, N. B. High thermal conductivity through simultaneously aligned polyethylene lamellae and graphene nanoplatelets. *Nanoscale* **2017**, *9* (35), 12867-12873.

(14) Shahil, K. M.; Balandin, A. A. Graphene–multilayer graphene nanocomposites as highly efficient thermal interface materials. *Nano letters* **2012**, *12* (2), 861-867.

(15) Xu, Y.; Wang, X.; Hao, Q. A mini review on thermally conductive polymers and polymer-based composites. *Composites Communications* **2021**, *24*, 100617.

(16) Lin, S.; Buehler, M. J. The effect of non-covalent functionalization on the thermal conductance of graphene/organic interfaces. *Nanotechnology* **2013**, *24* (16), 165702.

(17) Georgakilas, V.; Otyepka, M.; Bourlinos, A. B.; Chandra, V.; Kim, N.; Kemp, K. C.; Hobza, P.; Zboril, R.; Kim, K. S. Functionalization of graphene: covalent and non-covalent approaches, derivatives and applications. *Chemical reviews* **2012**, *112* (11), 6156-6214.

(18) Wang, M.; Hu, N.; Zhou, L.; Yan, C. Enhanced interfacial thermal transport across graphene–polymer interfaces by grafting polymer chains. *Carbon* **2015**, *85*, 414-421.

(19) Shen, X.; Wang, Z.; Wu, Y.; Liu, X.; He, Y.-B.; Kim, J.-K. Multilayer graphene enables higher efficiency in improving thermal conductivities of graphene/epoxy composites. *Nano letters* **2016**, *16* (6), 3585-3593.

(20) Miao, Z.; Li, X.; Zhi, L. Controlled functionalization of graphene with carboxyl moieties toward multiple applications. *RSC advances* **2016**, *6* (63), 58561-58565.

(21) Hummers, W. S.; Offeman, R. E. Preparation of Graphitic Oxide. *Journal of the American Chemical Society* **1958**, *80* (6), 1339-1339. DOI: 10.1021/ja01539a017.

(22) Wang, M.; Galpaya, D.; Lai, Z. B.; Xu, Y.; Yan, C. Surface functionalization on the thermal conductivity of graphene–polymer nanocomposites. *International Journal of Smart and Nano Materials* **2014**, *5* (2), 123-132.

(23) Konatham, D.; Striolo, A. Thermal boundary resistance at the graphene-oil interface. *Applied Physics Letters* **2009**, *95* (16), 163105.

(24) Ganguli, S.; Roy, A. K.; Anderson, D. P. Improved thermal conductivity for chemically functionalized exfoliated graphite/epoxy composites. *Carbon* **2008**, *46* (5), 806-817.

(25) Teng, C.-C.; Ma, C.-C. M.; Chiou, K.-C.; Lee, T.-M. Thermal conductivity and morphology of non-covalent functionalized graphene/epoxy nanocomposites. In *2010 5th International Microsystems Packaging Assembly and Circuits Technology Conference*, 2010; IEEE: pp 1-4.

(26) Yang, H.; Li, J.-S.; Zeng, X. Correlation between molecular structure and interfacial properties of edge or basal plane modified graphene oxide. *ACS Applied Nano Materials* **2018**, *1* (6), 2763-2773.

(27) Mungse, H. P.; Kumar, N.; Khatri, O. P. Synthesis, dispersion and lubrication potential of basal plane functionalized alkylated graphene nanosheets. *RSC advances* **2015**, *5* (32), 25565-25571.

(28) Rezvani Moghaddam, A.; Kamkar, M.; Ranjbar, Z.; Sundararaj, U.; Jannesari, A.; Ranjbar, B. Tuning the network structure of graphene/epoxy nanocomposites by controlling edge/basal localization of functional groups. *Industrial & Engineering Chemistry Research* **2019**, *58* (47), 21431-21440.

(29) Xiang, Z.; Dai, Q.; Chen, J. F.; Dai, L. Edge functionalization of graphene and two-dimensional covalent organic polymers for energy conversion and storage. *Advanced Materials* **2016**, *28* (29), 6253-6261.

(30) Pei, S.; Cheng, H.-M. The reduction of graphene oxide. *Carbon* **2012**, *50* (9), 3210-3228. Eigler, S.; Grimm, S.; Enzelberger-Heim, M.; Müller, P.; Hirsch, A. Graphene oxide: efficiency of reducing agents. *Chemical Communications* **2013**, *49* (67), 7391-7393.

(31) Compton, O. C.; Nguyen, S. T. Graphene oxide, highly reduced graphene oxide, and graphene: versatile building blocks for carbon-based materials. *small* **2010**, *6* (6), 711-723.

(32) Bagri, A.; Grantab, R.; Medhekar, N. V.; Shenoy, V. B. Stability and Formation Mechanisms of Carbonyl- and Hydroxyl-Decorated Holes in Graphene Oxide. *The Journal of Physical Chemistry C* **2010**, *114* (28), 12053-12061. DOI: 10.1021/jp908801c.

(33) Fugallo, G.; Cepellotti, A.; Paulatto, L.; Lazzeri, M.; Marzari, N.; Mauri, F. Thermal conductivity of graphene and graphite: collective excitations and mean free paths. *Nano letters* **2014**, *14* (11), 6109-6114.

(34) Mu, X.; Wu, X.; Zhang, T.; Go, D. B.; Luo, T. Thermal transport in graphene oxide–from ballistic extreme to amorphous limit. *Scientific reports* **2014**, *4* (1), 1-9.

(35) Shenogin, S.; Bodapati, A.; Xue, L.; Ozisik, R.; Keblinski, P. Effect of chemical functionalization on thermal transport of carbon nanotube composites. *Applied Physics Letters* **2004**, *85* (12), 2229-2231.

(36) Wang, G.; Yang, J.; Park, J.; Gou, X.; Wang, B.; Liu, H.; Yao, J. Facile synthesis and characterization of graphene nanosheets. *The Journal of Physical Chemistry C* **2008**, *112* (22), 8192-8195. Choi, Y. S.; Yeo, C.-s.; Kim, S. J.; Lee, J.-Y.; Kim, Y.; Cho, K. R.; Ju, S.; Hong, B. H.; Park, S. Y. Multifunctional reduced graphene oxide-CVD graphene core–shell fibers. *Nanoscale* **2019**, *11* (26), 12637-12642.

(37) Navaee, A.; Salimi, A. Efficient amine functionalization of graphene oxide through the Bucherer reaction: an extraordinary metal-free electrocatalyst for the oxygen reduction reaction. *Rsc Advances* **2015**, *5* (74), 59874-59880. Hussein, A.; Sarkar, S.; Oh, D.; Lee, K.; Kim, B. Epoxy/p-phenylenediamine functionalized graphene oxide composites and evaluation of their fracture toughness and tensile properties. *Journal of Applied Polymer Science* **2016**, *133* (34).

(38) Han, D.; Zhao, Y.-H.; Zhang, Y.-F.; Bai, S.-L. Vertically and compactly rolled-up reduced graphene oxide film/epoxy composites: a two-stage reduction method for graphene-based thermal interfacial materials. *RSC advances* **2015**, *5* (114), 94426-94435.

(39) Hwang, Y.; Heo, Y.; Yoo, Y.; Kim, J. The addition of functionalized graphene oxide to polyetherimide to improve its thermal conductivity and mechanical properties. *Polymers for Advanced Technologies* **2014**, *25* (10), 1155-1162.

(40) Bernal, M. M.; Di Pierro, A.; Novara, C.; Giorgis, F.; Mortazavi, B.; Saracco, G.; Fina, A. Edge-Grafted Molecular Junctions between Graphene Nanoplatelets: Applied Chemistry to Enhance Heat Transfer in Nanomaterials. *Advanced Functional Materials* **2018**, *28* (18), 1706954.

(41) Muthaiah, R.; Tarannum, F.; Danayat, S.; Annam, R. S.; Nayal, A. S.; Yedukondalu, N.; Garg, J. Superior effect of edge relative to basal plane functionalization of graphene in enhancing polymer-graphene nanocomposite thermal conductivity-A combined molecular dynamics and Greens functions study. *arXiv preprint arXiv:2201.01011* **2022**.

(42) Balandin, A. A. Thermal properties of graphene and nanostructured carbon materials. *Nature materials* **2011**, *10* (8), 569-581.

(43) Zhang, W.; Fisher, T. S.; Mingo, N. The Atomistic Green's Function Method: An Efficient Simulation Approach for Nanoscale Phonon Transport. *Numerical Heat Transfer, Part B: Fundamentals* **2007**, *51* (4), 333-349. DOI: 10.1080/10407790601144755.

(44) Yuge, R.; Zhang, M.; Tomonari, M.; Yoshitake, T.; Iijima, S.; Yudasaka, M. Site identification of carboxyl groups on graphene edges with Pt derivatives. *ACS nano* **2008**, *2* (9), 1865-1870.

(45) Radovic, L. R.; Mora-Vilches, C. V.; Salgado-Casanova, A. J.; Buljan, A. Graphene functionalization: Mechanism of carboxyl group formation. *Carbon* **2018**, *130*, 340-349.

(46) GRAPHENE SUPERMARKET. https://graphene-supermarket.com/Graphene-Nanopowder-AO-4-60-nm-Flakes-25-g.html.

(47) MilliporeSigma. https://www.sigmaaldrich.com/US/en/product/aldrich/700207/.

(48) Wu, H.; Drzal, L. T. High thermally conductive graphite nanoplatelet/polyetherimide composite by precoating: Effect of percolation and particle size. *Polymer composites* **2013**, *34* (12), 2148-2153.

(49) Marcano, D. C.; Kosynkin, D. V.; Berlin, J. M.; Sinitskii, A.; Sun, Z.; Slesarev, A.; Alemany, L. B.; Lu, W.; Tour, J. M. Improved synthesis of graphene oxide. *ACS nano* **2010**, *4* (8), 4806-4814. Yadav, N.; Lochab, B. A comparative study of graphene oxide: Hummers, intermediate and improved method. *FlatChem* **2019**, *13*, 40-49. Botas, C.; Álvarez, P.; Blanco, C.; Santamaría, R.; Granda, M.; Ares, P.; Rodríguez-Reinoso, F.; Menéndez, R. The effect of the parent graphite on the structure of graphene oxide. *Carbon* **2012**, *50* (1), 275-282.

(50) Bepete, G.; Pénicaud, A.; Drummond, C.; Anglaret, E. Raman Signatures of Single Layer Graphene Dispersed in Degassed Water,"'Eau de Graphene'". *The Journal of Physical Chemistry C* **2016**, *120* (49), 28204-28214.

(51) Ganguly, A.; Sharma, S.; Papakonstantinou, P.; Hamilton, J. Probing the thermal deoxygenation of graphene oxide using high-resolution in situ X-ray-based spectroscopies. *The Journal of Physical Chemistry C* **2011**, *115* (34), 17009-17019.

(52) Chen, Q.; Yan, X.; Wu, L.; Xiao, Y.; Wang, S.; Cheng, G.; Zheng, R.; Hao, Q. Small-Nanostructure-Size-Limited Phonon Transport within Composite Films Made of Single-Wall Carbon Nanotubes and Reduced Graphene Oxides. *ACS Applied Materials & Interfaces* **2021**, *13* (4), 5435-5444.

(53) Cao, H.; Wu, X.; Yin, G.; Warner, J. H. Synthesis of adenine-modified reduced graphene oxide nanosheets. *Inorganic chemistry* **2012**, *51* (5), 2954-2960.

(54) Li, J.; Zeng, X.; Ren, T.; Van der Heide, E. The preparation of graphene oxide and its derivatives and their application in bio-tribological systems. *Lubricants* **2014**, *2* (3), 137-161.

(55) Chhabra, V. A.; Deep, A.; Kaur, R.; Kumar, R. Functionalization of graphene using carboxylation process. *Int. J. Sci. Emerg. Technol* **2012**, *4*, 13-19.

(56) Krishnamoorthy, K.; Veerapandian, M.; Yun, K.; Kim, S.-J. The chemical and structural analysis of graphene oxide with different degrees of oxidation. *Carbon* **2013**, *53*, 38-49.

(57) Shahriary, L.; Athawale, A. A. Graphene oxide synthesized by using modified hummers approach. *Int. J. Renew. Energy Environ. Eng* **2014**, *2* (01), 58-63.

(58) Lerf, A.; He, H.; Forster, M.; Klinowski, J. Structure of graphite oxide revisited. *The Journal of Physical Chemistry B* **1998**, *102* (23), 4477-4482.

(59) Tabish, T. A.; Pranjol, M. Z. I.; Horsell, D. W.; Rahat, A. A.; Whatmore, J. L.; Winyard, P. G.; Zhang, S. Graphene oxide-based targeting of extracellular cathepsin D and cathepsin L as a novel anti-metastatic enzyme cancer therapy. *Cancers* **2019**, *11* (3), 319.

(60) Park, J.; Kim, Y. S.; Sung, S. J.; Kim, T.; Park, C. R. Highly dispersible edge-selectively oxidized graphene with improved electrical performance. *Nanoscale* **2017**, *9* (4), 1699-1708.

(61) Hack, R.; Correia, C. H. G.; Zanon, R. A. d. S.; Pezzin, S. H. Characterization of graphene nanosheets obtained by a modified Hummer's method. *Matéria (Rio de Janeiro)* **2018**, *23* (1).

(62) Aliyev, E.; Filiz, V.; Khan, M. M.; Lee, Y. J.; Abetz, C.; Abetz, V. Structural characterization of graphene oxide: Surface functional groups and fractionated oxidative debris. *Nanomaterials* **2019**, *9* (8), 1180.

(63) Stobinski, L.; Lesiak, B.; Malolepszy, A.; Mazurkiewicz, M.; Mierzwa, B.; Zemek, J.; Jiricek, P.; Bieloshapka, I. Graphene oxide and reduced graphene oxide studied by the XRD, TEM and electron spectroscopy methods. *Journal of Electron Spectroscopy and Related Phenomena* **2014**, *195*, 145-154.

(64) Drewniak, S.; Muzyka, R.; Stolarczyk, A.; Pustelny, T.; Kotyczka-Morańska, M.; Setkiewicz, M. Studies of reduced graphene oxide and graphite oxide in the aspect of their possible application in gas sensors. *Sensors* **2016**, *16* (1), 103.

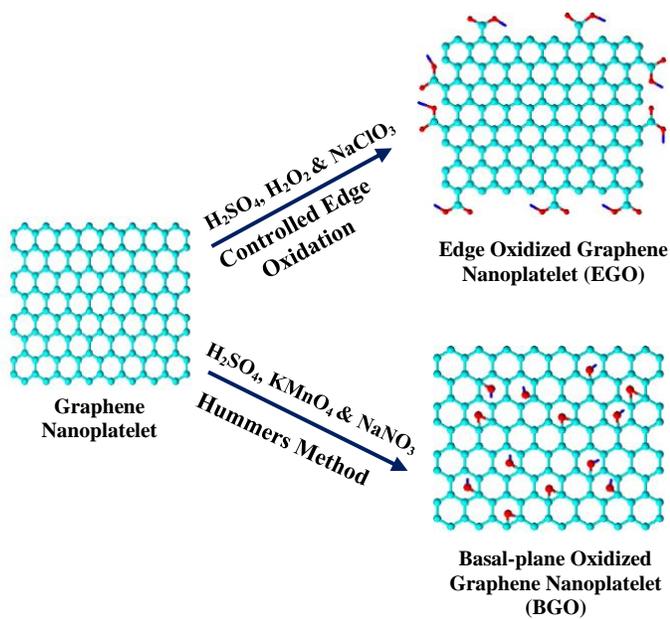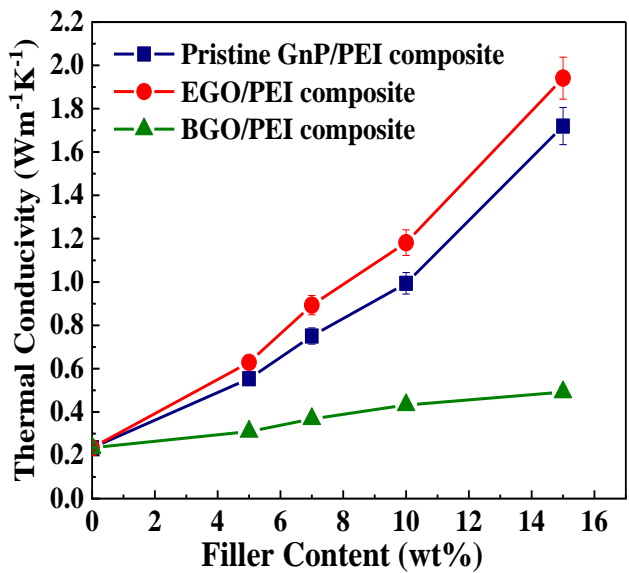